\documentclass[12pt]{article}
\usepackage{epsf}
\title{ {\bf
CP violation in the lepton flavor violating interactions 
$\mu\rightarrow e\gamma$ and $\tau\rightarrow \mu\gamma$ }}
\author{\vspace{1cm}\\
        \thanks{E-mail address:
        eiltan@heraklit.physics.metu.edu.tr}
 \\
        Physics Department, Middle East Technical University \\
        Ankara, Turkey\\}

\date{}

\begin{document}
\setlength{\baselineskip}{24pt}
\maketitle
\setlength{\baselineskip}{7mm}
\begin{abstract}
We calculate the possible CP violating asymmetries for LFV decays 
$\mu\rightarrow e\gamma$ and $\tau\rightarrow \mu\gamma$. Our predictions
depend on the chosen new model independent contribution. The result of
measurements of such asymmetries for these decays will open a new window to
determine the physics beyond the SM.    
\end{abstract} 
\thispagestyle{empty}
\newpage
\setcounter{page}{1}
%%%
%%%
\section{Introduction}
Lepton Flavor Violating (LFV) interactions are among the most promising
candidates to understand the physics beyond the Standard model (SM). The 
improvement of their experimental measurements forces to make more elaborate 
theoretical calculations and to determine the unknown parameters existing in
the models used. $\mu\rightarrow e\gamma$ and $\tau\rightarrow \mu\gamma$
decays are the examples for the LFV decays and the current limits for their 
branching ratios ($BR$) are $1.2\times 10^{-11}$ \cite{Brooks} and  
$1.1\times 10^{-6}$ \cite{Ahmed} respectively. 

LFV interactions have been analysed in different models in the literature. 
They were studied in a model independent way in \cite{Chang}, in the 
framework of model III 2HDM \cite{Diaz}, in supersymmetric models 
\cite{Barbieri1,Barbieri2,Barbieri3,Ciafaloni,Duong,Couture,Okada}. 
Recently, the electromagnetic suppression of the decay rate of $\mu
\rightarrow e\gamma$ has been predicted in \cite{Czarnecki}. Further, the 
processes $\tau\rightarrow \mu\gamma$ and $\mu\rightarrow e\gamma$ have 
been studied as probes of neutrino mass models in \cite{Lavignac}.    

LFV processes do not exist in the Standard model (SM) if the Cabibbo-
Kobayashi-Maskawa (CKM) type matrix in the leptonic sector vanishes and 
this stimulates one to go beyond.  The general two Higgs doublet model, 
so called model III, permits such interactions which appear at least in the 
loop level, with the internal mediating neutral Higgs bosons $h_0$ and $A_0$.
Note that, in this case, there is no charged Flavor Changing (FC) 
interaction. There are large number of  free parameters and their strength 
can be detemined by the experimental data. Choice of complex Yukawa couplings
causes the CP violating effects which can also provide comprehensive 
information in the determination of free parameters of the various 
theoretical models. Non-zero Electric Dipole Moments (EDM) of the elementary 
particles are sign of such violations.

In this work, we study the possiblity of CP asymmetry $A_{CP}$ of decays 
$\mu\rightarrow e \gamma$ and $\tau\rightarrow \mu \gamma$ in the model III. 
Even if the Yukawa couplings are taken as complex in general, the model III
does not ensure CP asymmetry for the decays under consideration. However,
by correcting the decay rates of these processes with the additional complex 
contribution, which may come from the physics beyond the model III, a
measurable $A_{CP}$ can be obtained. Here, we assume that the complexity of
the new contribution is due to not the Yukawa type couplings, but probably
to radiative corrections in this model. Therefore, the forthcoming 
experimental measurements of possible $A_{CP}$ for both processes open a 
new window understand the physics beyond the SM.

The paper is organized as follows:
In Section 2, we present the possible CP violating asymmetry for LFV decays 
$\mu\rightarrow e\gamma$ and $\tau\rightarrow \mu\gamma$.
Section 3 is devoted to discussion and our conclusions.
%%%
%%%
\section{CP violation in LFV interactions $\mu\rightarrow e\gamma$ and 
$\tau\rightarrow \mu\gamma$ } 
Our starting point is type III 2HDM which permits flavor changing neutral 
currents (FCNC) at tree level. The  Yukawa interaction for the leptonic 
sector in the model III is
\begin{eqnarray}
{\cal{L}}_{Y}=
\eta^{E}_{ij} \bar{l}_{i L} \phi_{1} E_{j R}+
\xi^{E}_{ij} \bar{l}_{i L} \phi_{2} E_{j R} + h.c. \,\,\, ,
\label{lagrangian}
\end{eqnarray}
where $i,j$ are family indices of leptons, $L$ and $R$ denote chiral 
projections $L(R)=1/2(1\mp \gamma_5)$, $\phi_{i}$ for $i=1,2$, are the 
two scalar doublets, $l_{i L}$ and $E_{j R}$ are lepton doublets and
singlets respectively. 
Here $\phi_{1}$ and $\phi_{2}$ are chosen as
\begin{eqnarray}
\phi_{1}=\frac{1}{\sqrt{2}}\left[\left(\begin{array}{c c} 
0\\v+H^{0}\end{array}\right)\; + \left(\begin{array}{c c} 
\sqrt{2} \chi^{+}\\ i \chi^{0}\end{array}\right) \right]\, ; 
\phi_{2}=\frac{1}{\sqrt{2}}\left(\begin{array}{c c} 
\sqrt{2} H^{+}\\ H_1+i H_2 \end{array}\right) \,\, ,
\label{choice}
\end{eqnarray}
and the vacuum expectation values are  
\begin{eqnarray}
<\phi_{1}>=\frac{1}{\sqrt{2}}\left(\begin{array}{c c} 
0\\v\end{array}\right) \,  \, ; 
<\phi_{2}>=0 \,\, .
\label{choice2}
\end{eqnarray}
With this choice, the SM particles can be collected in the first doublet 
and the new particles in the second one. Here the bosons $H_1$ and $H_2$ 
are the neutral CP even $h^0$ and CP odd $A^0$, respectively since there is
no mixing between CP even neutral Higgs bosons $H^0$ and $h^0$ at tree level.
The part which produce FCNC (at tree level) is  
\begin{eqnarray}
{\cal{L}}_{Y,FC}=
\xi^{E}_{ij} \bar{l}_{i L} \phi_{2} E_{j R} + h.c. \,\, .
\label{lagrangianFC}
\end{eqnarray}
The Yukawa matrices $\xi^{E}_{ij}$ have in general complex entries. 
Note that in the following we replace $\xi^{E}$ with $\xi^{E}_{N}$ where 
"N" denotes the word "neutral". 

\begin{figure}[htb]
\vskip -0.0truein
\centering
\epsfxsize=6.0in
\leavevmode\epsffile{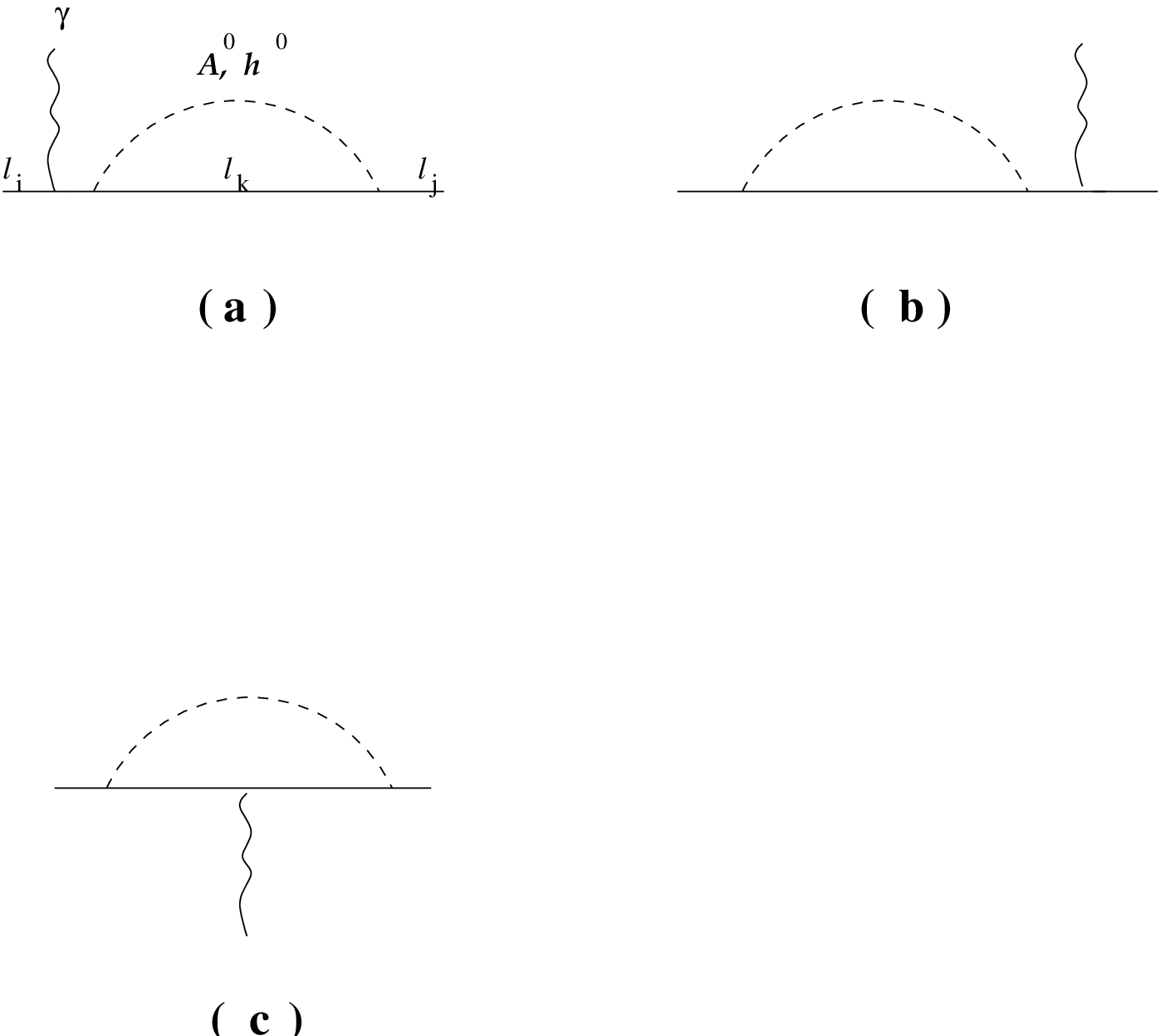}
\vskip 0.5truein
\caption[]{One loop diagrams contribute to the LFV decays $l_i\rightarrow
l_j\gamma$ with $i\neq j$. Solid line corresponds to the lepton, curly line 
to the electromagnetic field, dashed line to the fields $h_0$ and $A_0$. 
Here $l_k$ denotes the leptons $e,\mu,\tau$.}
\label{fig1}
\end{figure}

Now, let us consider the lepton number violating process $\mu\rightarrow e 
\gamma$. Here, we expect that the main contribution to this decay comes from 
the neutral Higgs bosons, namely, $h_0$ and $A_0$, in the leptonic 
sector of the model III, (see Fig. \ref{fig1}). Using on-shell renormalization 
scheme the self energy diagrams vanish and only the vertex diagram (Fig. 
\ref{fig1}-c) contributes. By taking $\tau$ lepton for the internal 
line, the decay width $\Gamma$ becomes \cite{IltanLFV} 
\begin{eqnarray}
\Gamma (\mu\rightarrow e\gamma)=c_1(|A_1|^2+|A_2|^2)\,\, ,
\label{DWmuegam}
\end{eqnarray}
where
\begin{eqnarray}
A_1&=&Q_{\tau} \frac{1}{8\,m_{\mu}\,m_{\tau}} \bar{\xi}^E_{N,\tau e}\, 
\bar{\xi}^E_{N,\tau \mu}\, (F_1 (y_{h_0})-F_1 (y_{A_0}))
\nonumber \,\, , \\
A_2&=&Q_{\tau} \frac{1}{8\,m_{\mu}\,m_{\tau}} \bar{\xi}^{E *}_{N,e \tau}\, 
\bar{\xi}^{E *}_{N,\mu \tau}\, (F_1 (y_{h_0})-F_1 (y_{A_0})) \,\, ,
\label{A1A2}
\end{eqnarray}
$c_1=\frac{G_F^2 \alpha_{em} m^5_{\mu}}{32 \pi^4}$ and the function 
$F_1 (w)$ reads 
\begin{eqnarray}
F_1 (w)&=&\frac{w\,(3-4\,w+w^2+2\,ln\,w)}{(-1+w)^3} , 
\label{functions1}
\end{eqnarray}
with $y_{H}=\frac{m^2_{\tau}}{m^2_{H}}$.  In eq. (\ref{A1A2}),  
$\bar{\xi}^{E}_{N,ij}$ is defined as $\xi^{E}_{N,ij}=
\sqrt{\frac{4\,G_F}{\sqrt{2}}}\, \bar{\xi}^{E}_{N,ij}$ and the amplitudes 
$A_1$ and $A_2$ have right and left chirality  respectively. In our
calculations we ignore the contributions coming from internal $\mu$ and 
$e$ leptons, respecting our assumption on the Yukawa couplings 
(see Discussion). 

At this stage, we calculate the CP asymmetry $A_{CP}$ of the 
process $\mu \rightarrow e \gamma$, given by the expression 
\begin{eqnarray}
A_{CP}=\frac{\Gamma - \bar{\Gamma}}{\Gamma + \bar{\Gamma}}
\label{ACP1}
\end{eqnarray}
where $\bar{\Gamma}$ is the decay width for the CP conjugate process. 
However, in the framework of the model III, $A_{CP}$ vanishes and one needs 
an extra quantity to switch on the CP asymmetry. Therefore, we assume that 
there is an additional small and complex contribution $\chi$ due to the 
physics beyond the model III such that the factor $\bar{\xi}^E_{N,\tau e}
\, \bar{\xi}^E_{N,\tau \mu}\, (F_1 (y_{h_0})-F_1 (y_{A_0}))$ is corrected as 
\begin{eqnarray}
\bar{\xi}^E_{N,\tau e}\, \bar{\xi}^E_{N,\tau \mu}\, (F_1 (y_{h_0})-
F_1 (y_{A_0}))+\chi \nonumber \, .
\end{eqnarray}
Further, we force that the complexity of $\chi$ comes from the possible 
radiative corrections but not from the Yukawa type couplings, in the model 
beyond the model III. This choice of $\chi$ brings a non-zero CP violation 
for the process $\mu\rightarrow e \gamma$ :
\begin{eqnarray}
A_{CP}=2\, \frac{|\bar{\xi}^E_{N,\tau e}\, \bar{\xi}^E_{N,\tau \mu}|\, 
(F_1 (y_{h_0})-F_1 (y_{A_0}))\, |\chi|\, sin\,\theta_{\chi}\,
sin\,(\theta_{\tau e}+\theta_{\tau \mu})}{\Phi}
\label{ACP2}
\end{eqnarray}
where
\begin{eqnarray}
\Phi&=& |\bar{\xi}^E_{N,\tau e}\, 
\bar{\xi}^E_{N,\tau \mu}|^2\, (F_1 (y_{h_0})-F_1 (y_{A_0}))^2+|\chi|^2 
\nonumber \\ 
&+& 2\,|\bar{\xi}^E_{N,\tau e}\, \bar{\xi}^E_{N,\tau \mu}|\, 
(F_1 (y_{h_0})-F_1 (y_{A_0}))\, |\chi|\, cos\,\theta_{\chi}\,
cos\,(\theta_{\tau e}+\theta_{\tau \mu})
\label{ACP2a}
\end{eqnarray}
with $\chi=e^{i\,\theta_{\chi}}\,|\chi|$, $\bar{\xi}^E_{N,\tau e}=
e^{i\,\theta_{\tau e}}\,|\bar{\xi}^E_{N,\tau e}|$ and $\bar{\xi}^E_{N,\tau
\mu}= e^{i\,\theta_{\tau \mu}}\,|\bar{\xi}^E_{N,\tau \mu}|$.
Of course, the amount of $A_{CP}$ produced depends on the amount of new 
quantity introduced, however even a small contribution may bring a measurable
$A_{CP}$ for this process.   

Now, we would like to discuss the similar analysis for another LFV decay, 
$\tau\rightarrow \mu\gamma$. The decay width for this process is calculated 
by taking only the $\tau$-lepton as an internal one \cite{IltanLFV} and it 
reads as
\begin{eqnarray}
\Gamma (\tau\rightarrow \mu\gamma)=c_2(|B_1|^2+|B_2|^2)\,\, ,
\end{eqnarray}
\label{DWtaumugam}
where
\begin{eqnarray}
B_1&=& Q_{\tau} \frac{1}{48\,m_{\mu}\,m_{\tau}} \bar{\xi}^E_{N,\tau \mu} 
\Big{\{}\bar{\xi}^{E *}_{N,\tau \tau} (G_1 (y_{h_0})+G_1 (y_{A_0}))+ 
6 \bar{\xi}^{E}_{N,\tau \tau} (F_1 (y_{h_0})-F_1 (y_{A_0})\Big{\}} 
\nonumber \,\, , \\
B_2&=& Q_{\tau} \frac{1}{48\,m_{\mu}\,m_{\tau}} \bar{\xi}^{E *}_{N,\mu\tau} 
\Big{\{}\bar{\xi}^{E}_{N,\tau \tau} (G_1 (y_{h_0})+G_1 (y_{A_0}))+ 
6 \bar{\xi}^{E *}_{N,\tau \tau} (F_1 (y_{h_0})-F_1 (y_{A_0})\Big{\}} 
\,\, ,
\label{B1B2}
\end{eqnarray}
and $c_2=\frac{G_F^2 \alpha_{em} m^5_{\tau}}{32 \pi^4}$. Here the amplitudes 
$B_1$ and $B_2$ have right and left chirality, respectively. The function 
$F_1 (w)$ is given in eq. (\ref{functions1}) and $G_1 (w)$ is  
\begin{eqnarray}
G_1 (w)&=&\frac{w\,(2+3\,w-6\,w^2+w^3+ 6\,w\,ln\,w)}{(-1+w)^4}
\nonumber \,\, . 
\label{functions2}
\end{eqnarray}

$A_{CP}$ in this process vanishes in the model III, similar to the one in
the decay $\mu \rightarrow e\gamma$ and we will follow the same procedure
given above. By correcting the combination  $\bar{\xi}^E_{N,\tau \mu} 
\Big ( \bar{\xi}^{E *}_{N,\tau \tau} (G_1 (y_{h_0})+G_1 (y_{A_0}))+ 
6\, \bar{\xi}^{E}_{N,\tau \tau} (F_1 (y_{h_0})-F_1 (y_{A_0}) \Big )$ with 
the additional small and complex  quantity $\rho$ due to the physics 
beyond the model III as 
\begin{eqnarray}
\bar{\xi}^E_{N,\tau \mu} 
\Big ( \bar{\xi}^{E *}_{N,\tau \tau} (G_1 (y_{h_0})+G_1 (y_{A_0}))+ 
6\, \bar{\xi}^{E}_{N,\tau \tau} (F_1 (y_{h_0})-F_1 (y_{A_0}) \Big )+\rho 
\nonumber \,\, , 
\end{eqnarray}
we get 
\begin{eqnarray}
A_{CP}=\frac{\Lambda}{\Omega}
\label{ACP3}
\end{eqnarray}
where 
\begin{eqnarray}
\Lambda &=& 2\, |\bar{\xi}^E_{N,\tau \mu}\, \bar{\xi}^{E *}_{N,\tau
\tau}|\,|\rho|\,sin\,\theta_{\rho} \{ sin\,(\theta_{\tau\mu}-
\theta_{\tau\tau}) (G_1 (y_{h_0})+G_1 (y_{A_0}))
\nonumber \\ &+& 
6\, sin\,(\theta_{\tau\mu}+ \theta_{\tau\tau}) (F_1 (y_{h_0})-F_1 (y_{A_0}))
\} \, ,
\label{ACP3a}
\end{eqnarray}
and 
\begin{eqnarray}
\Omega &=& |\bar{\xi}^E_{N,\tau \mu}\, \bar{\xi}^{E *}_{N,\tau\tau}|^2\,
\Bigg( (G_1 (y_{h_0})+G_1 (y_{A_0}))^2+ 36\, (F_1 (y_{h_0})-F_1 (y_{A_0}))^2 
\Bigg) + |\rho|^2 \nonumber \\ &+& 
12\, |\bar{\xi}^E_{N,\tau \mu}\, \bar{\xi}^{E *}_{N,\tau\tau}|^2\,
(F_1 (y_{h_0})-F_1 (y_{A_0}))\, (G_1 (y_{h_0})+G_1 (y_{A_0}))\,
cos\,2\,\theta_{\tau\tau} 
\nonumber \\ &+&
2\, |\bar{\xi}^E_{N,\tau \mu}\, \bar{\xi}^{E *}_{N,\tau\tau}|\,|\rho|\, 
(G_1 (y_{h_0})+G_1 (y_{A_0}))\,cos\theta_{\rho}\,
cos\,(\theta_{\tau\mu}-\theta_{\tau\tau}) \nonumber \\ &+&
12\, |\bar{\xi}^E_{N,\tau \mu}\, \bar{\xi}^{E *}_{N,\tau\tau}|\,|\rho|\, 
(F_1 (y_{h_0})+F_1 (y_{A_0}))\,cos\theta_{\rho}\,
cos\,(\theta_{\tau\mu}+\theta_{\tau\tau}) \, ,
\label{ACP4}
\end{eqnarray}
with $\rho=e^{i\,\theta_{\rho}}\,|\rho|$ and  $\bar{\xi}^E_{N,\tau \tau}
=e^{i\,\theta_{\tau \tau}}\,|\bar{\xi}^E_{N,\tau \tau}|$.
Similar to the process $\mu\rightarrow e\gamma$, the amount of $A_{CP}$ 
strongly depends on the new quantity introduced and a small contribution 
may bring measurable $A_{CP}$ for this process also.  

%
%%%
%%%
\section{Discussion}
In the case of vanishing charged interactions, with the assumption that CKM 
type matrix in the leptonic sector does not exist, LFV interactions are 
possible in the one-loop, due to neutral Higgs bosons $h^0$ and $A^0$, in 
the framework of model III. In general, the Yukawa couplings 
$\bar{\xi}^E_{N,ij}, i,j=e, \mu, \tau$ appearing in the expressions are 
complex and they ensure non-zero lepton EDM. However, this scenario is not 
enough to get a CP violating asymmetry in the LFV processes, 
$\mu\rightarrow e \gamma$ and $\tau\rightarrow \mu \gamma$. To obtain such 
asymmetry, we introduce a small model independent correction term to the 
decay width with the following features:
\begin{itemize}
\item this term is due to the physics beyond the model III
\item it is complex and the complexity does not come from the Yukawa type
couplings 
\end{itemize}
The additional contributions respecting the above conditions bring non-zero
$A_{CP}$ to both processes underconsideration. However, this extra quantity
is completely unknown and the number of parameters, namely complex Yukawa 
couplings and new model independent quantity, increases in the numerical
calculations. To solve this problem, first, we assume that the Yukawa 
couplings $\bar{\xi}^{E}_{N,ij},\, i,j=e,\mu $, are small compared to 
$\bar{\xi}^{E}_{N,\tau\, i}\, i=e,\mu,\tau$ since the strength of them 
are related with the masses of leptons denoted by the indices of 
them, similar to the Cheng-Sher scenerio \cite{Sher}. Further, we take 
$\bar{\xi}^{E}_{N,ij}$  symmetric with respect to the indices $i$ 
and $j$. Therefore only the combination $\bar{\xi}^{E}_{N,\tau\, \mu} 
\bar{\xi}^{E}_{N,\tau\, e}$ (the couplings $\bar{\xi}^{E}_{N,\tau\,
\tau}$ and  $\bar{\xi}^{E}_{N,\tau\,\mu}$) for the process 
$\mu\rightarrow e \gamma$ ($\tau\rightarrow \mu \gamma$) plays the main 
role in our analysis. $\bar{\xi}^{E}_{N,\tau\, \mu} \bar{\xi}^{E}_{N,\tau\, 
e}$ can be restricted using the experimental upper limit of the $BR$ of 
the process $\mu\rightarrow e \gamma$ \cite{Brooks}, 
\begin{eqnarray}
BR (\mu\rightarrow e \gamma) < 1.2\times 10^{-11}\, .
\label{muegamma}
\end{eqnarray}
(see \cite{IltanLFV} for details).
Here, we do not take the contribution of additional part coming from the
physics beyond model III since we assume that its effect on the 
constraint region is sufficiently small. Note that we take the additional
contribution $|\chi|$ as two orders smaller compared to 
$\bar{\xi}^E_{N,\tau e}\, \bar{\xi}^E_{N,\tau \mu}\, (F_1 (y_{h_0})-
F_1 (y_{A_0}))$. 

For the process $\tau\rightarrow \mu \gamma$, the coupling 
$\bar{\xi}^{E}_{N,\tau\, \mu}$ is restricted using the experimental upper 
and lower limits of $\mu$-lepton EDM (\cite {Abdullah}) 
\begin{eqnarray}
0.3\times 10^{-19}\, e-cm < d_{\mu} < 7.1\times 10^{-19}\, e-cm 
\label{muedmex}
\end{eqnarray}
(see \cite{IltanLFV} for details) and we do not use any constraint for the 
coupling $\bar{\xi}^{E}_{N,\tau\, \tau}$. For this decay, the additional
contribution $|\rho|$ is taken as two orders smaller compared to 
$\bar{\xi}^E_{N,\tau \mu} \Big ( \bar{\xi}^{E *}_{N,\tau \tau} (G_1 (y_{h_0})
+G_1 (y_{A_0}))+ 6\, \bar{\xi}^{E}_{N,\tau \tau} (F_1 (y_{h_0})-F_1 (y_{A_0}) 
\Big )$, similar to the previous process.

\begin{figure}[htb]
\vskip -3.0truein
\centering
\epsfxsize=6.8in
\leavevmode\epsffile{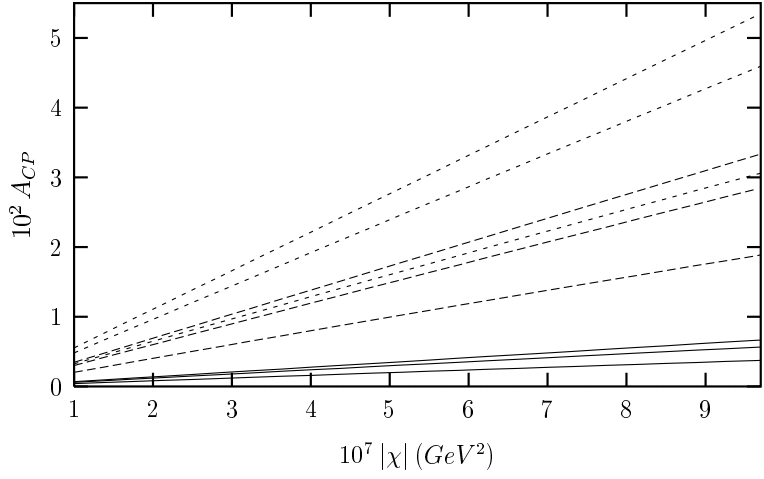}
\vskip -3.0truein
\caption[]{$A_{CP}$ of the process $\mu\rightarrow e\gamma$ as a function of 
$|\chi|$ for $sin\theta_{\tau\mu}=0.5$, $m_{h^0}=85\, GeV$ and $m_{A^0}=95
\, GeV$. Here, the solid lines show the case where $sin\,\theta_{\chi}=0.1$ 
and they correspond to $sin\theta_{\tau e}$ values $0.1$, $0.5$ and $0.8$ 
in the order from the lower one to the upper. Similarly, the dashed 
(small dashed) lines represent the case where $sin\,\theta_{\chi}=0.5$ 
($sin\,\theta_{\chi}=0.8$).}
\label{ACPalfmueg}
\end{figure}

Fig. \ref{ACPalfmueg} represents the new quantity $|\chi|$ dependence of 
$A_{CP}$ for $sin\theta_{\tau\mu}=0.5$, $m_{h^0}=85\, GeV$ and $m_{A^0}=95
\, GeV$. Here the solid lines show the case where $sin\,\theta_{\chi}=0.1$ 
and they correspond to $sin\theta_{\tau e}$ values $0.1$, $0.5$ and $0.8$ 
in the order from the lower one to the upper.  Similarly the dashed (small 
dashed) lines represents the case where $sin\,\theta_{\chi}=0.5$ 
($sin\,\theta_{\chi}=0.8$). Choosing $|\chi|$ at the order of $10^{-7} 
(GeV^2)$, $A_{CP}$ for the process can be at the order of the magnitude 
$10^{-2}$. As shown in this figure, $A_{CP}$ is sensitive to the CP parameters 
$sin\theta_{\tau\mu}$, $sin\theta_{\tau e}$ and obviously to 
$sin\,\theta_{\chi}$.
 
\begin{figure}[htb]
\vskip -3.0truein
\centering
\epsfxsize=6.8in
\leavevmode\epsffile{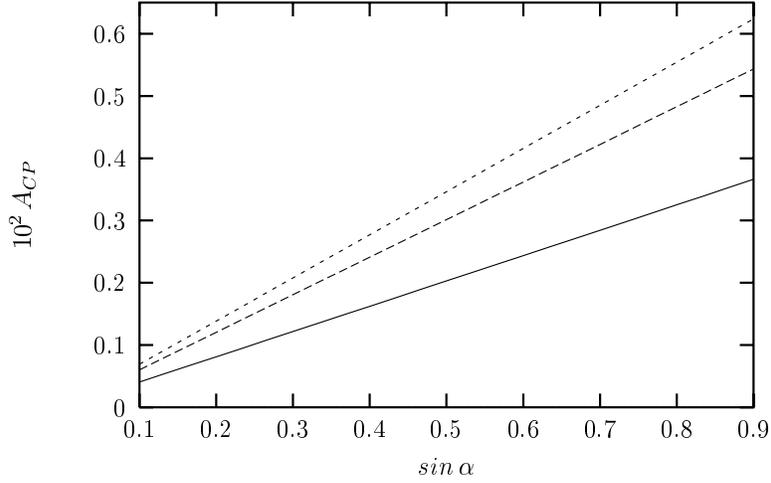}
\vskip -3.0truein
\caption[]{$A_{CP}$ of the process $\mu\rightarrow e\gamma$ as a function 
of $sin\,\theta_{\chi}$ for $|\chi|=10^{-7} (GeV^2)$, $sin\theta_{\tau e}
=0.5$, $m_{h^0}=85\, GeV$ and $m_{A^0}=95\, GeV$. Here, the solid line 
coresponds to $sin\, \theta_{\tau \mu}=0.1$, dashed line to $sin\, 
\theta_{\tau \mu}=0.5$ and small dashed line to $sin\, \theta_{\tau \mu}
=0.8$.}
\label{ACPsinalfmuegalfextrm7}
\end{figure}

In Fig. \ref{ACPsinalfmuegalfextrm7} we present $sin\,\theta_{\chi}$ 
dependence of $A_{CP}$ for $|\chi|=10^{-7} (GeV^2)$, $sin\theta_{\tau e}=
0.5$, $m_{h^0}=85\, GeV$ and $m_{A^0}=95\, GeV$. Here, the solid line
coresponds to $sin\, \theta_{\tau \mu}=0.1$, dashed line to 
$sin\, \theta_{\tau \mu}=0.5$ and small dashed line to $sin\, 
\theta_{\tau \mu}=0.8$. $A_{CP}$ increases with increasing values of the
parameters $sin\,\theta_{\chi}$, $sin\, \theta_{\tau e}$ and 
$sin\, \theta_{\tau \mu}$.

\begin{figure}[htb]
\vskip -3.0truein
\centering
\epsfxsize=6.8in
\leavevmode\epsffile{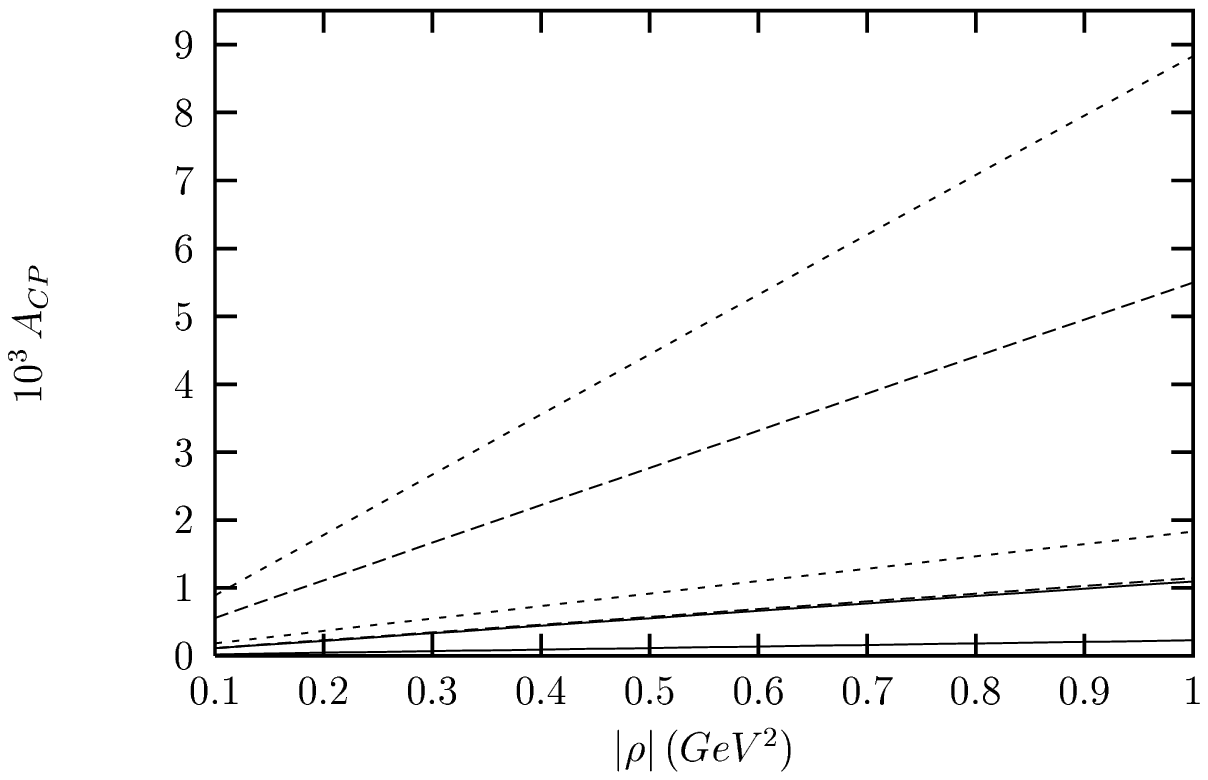}
\vskip -3.0truein
\caption[]{$A_{CP}$ of the process $\tau\rightarrow \mu\gamma$ as a function 
of $|\rho|$ for $\bar{\xi}^E_{N,\tau\tau}=100\, GeV$, $sin\theta_{\rho}
=0.1$, $sin\theta_{\tau\mu}=0.5$,  $m_{h^0}=85\, GeV$ and $m_{A^0}=95\, GeV$. 
Here, $A_{CP}$ is restricted between solid lines for $sin\theta_{\tau\tau}=
0.1$, dashed lines for $sin\theta_{\tau\tau}=0.5$ and small dashed lines 
for $sin\theta_{\tau\tau}=0.8$.}
\label{ACPalftaumugsinalf01}
\end{figure}

\begin{figure}[htb]
\vskip -3.0truein
\centering
\epsfxsize=6.8in
\leavevmode\epsffile{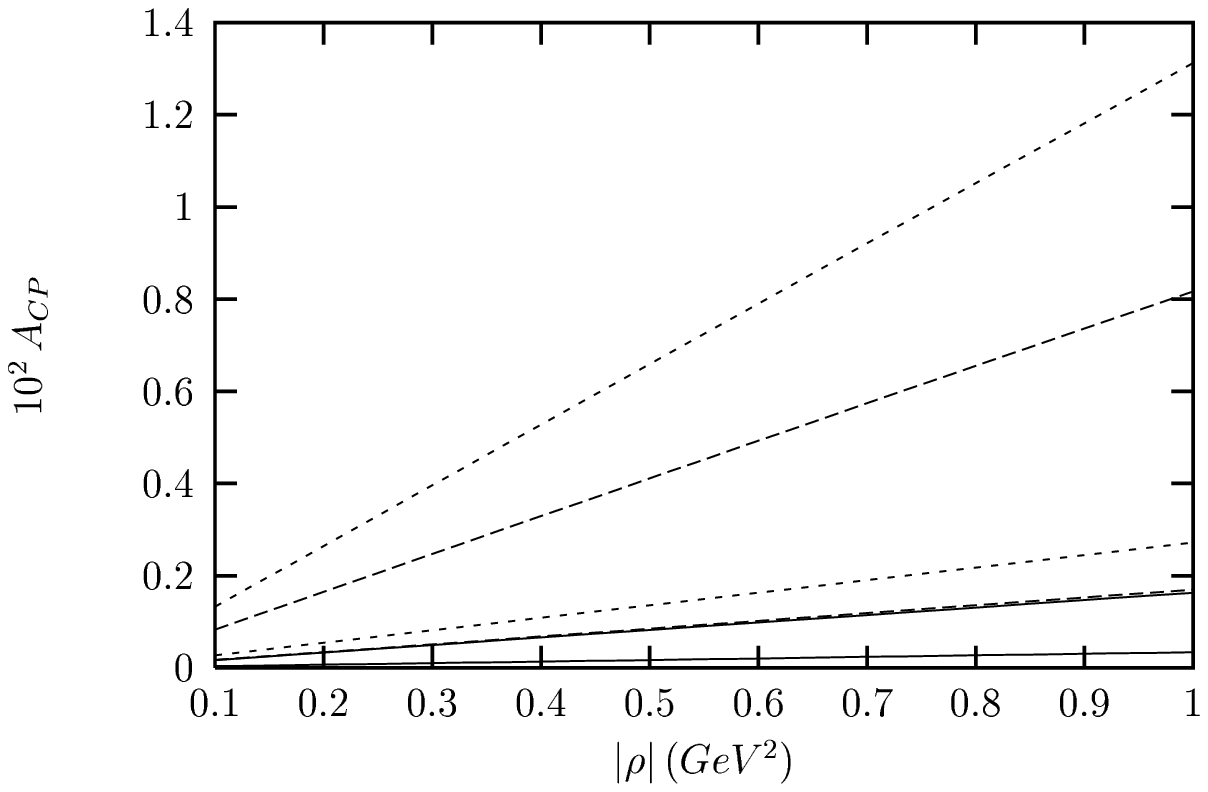}
\vskip -3.0truein
\caption[]{The same as Fig. \ref{ACPalftaumugsinalf01} but for 
$sin\theta_{\rho}=0.5$.}
\label{ACPalftaumugsinalf05}
\end{figure}

\begin{figure}[htb]
\vskip -3.0truein
\centering
\epsfxsize=6.8in
\leavevmode\epsffile{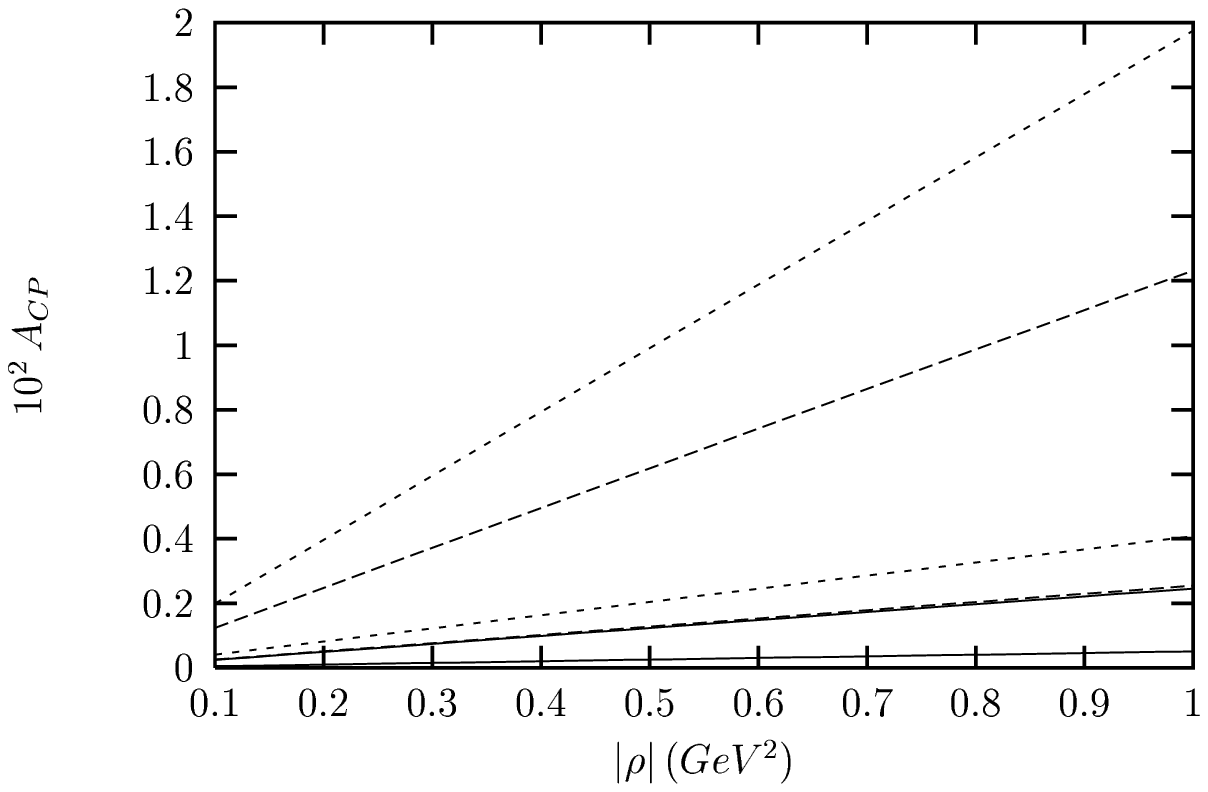}
\vskip -3.0truein
\caption[]{The same as Fig. \ref{ACPalftaumugsinalf01} but for 
$sin\theta_{\rho}=0.8$.}
\label{ACPalftaumugsinalf08}
\end{figure}

\begin{figure}[htb]
\vskip -3.0truein
\centering
\epsfxsize=6.8in
\leavevmode\epsffile{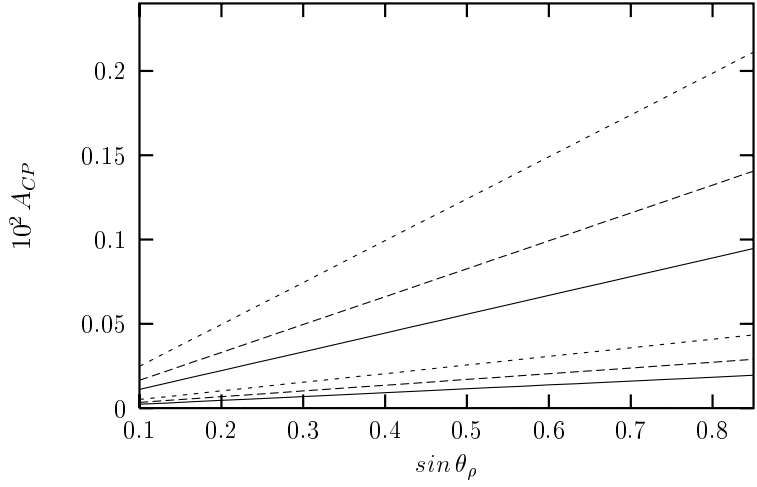}
\vskip -3.0truein
\caption[]{$A_{CP}$ of the process $\tau\rightarrow \mu\gamma$ as a function 
of $\sin\,\theta_{\rho}$ for $|\rho|=0.1\, (GeV^2)$, $sin\theta_{\tau \mu}=
0.5$, $m_{h^0}=85\, GeV$ and $m_{A^0}=95\, GeV$. Here, $A_{CP}$ is restricted 
between solid lines for $sin\theta_{\tau\tau}=0.1$, dashed lines for 
$sin\theta_{\tau\tau}=0.5$ and small dashed lines for $sin\theta_{\tau\tau}
=0.8$.}
\label{ACPsinalftaumugalfextrm1}
\end{figure}

\begin{figure}[htb]
\vskip -3.0truein
\centering
\epsfxsize=6.8in
\leavevmode\epsffile{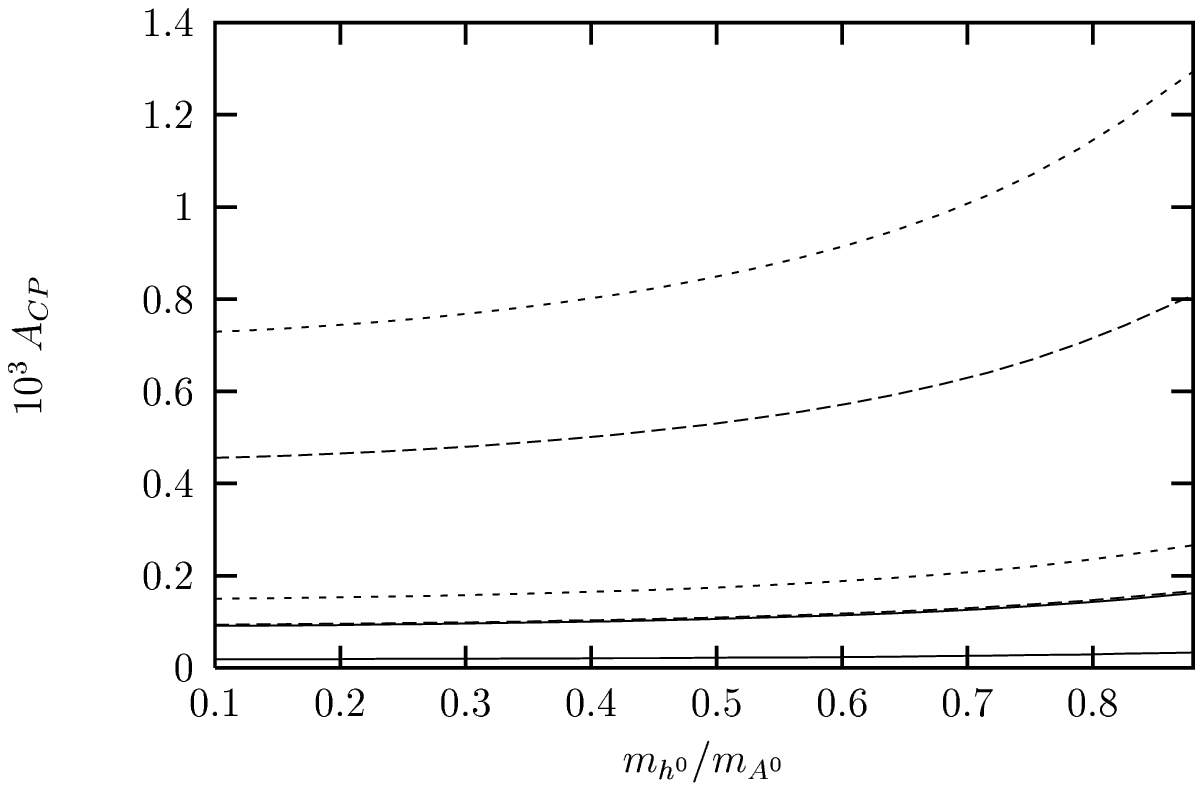}
\vskip -3.0truein
\caption[]{$A_{CP}$ of the process $\tau\rightarrow \mu\gamma$ as a 
function of the mass ratio $m_{h^0}/m_{A^0}$ for 
$\bar{\xi}^E_{\tau\tau}=100\, GeV$, $sin\theta_{\tau \mu}
=sin\theta_{\tau \tau}=0.5$, $|\rho|=0.1\, (GeV^2)$ and $m_{h^0}=85\, GeV$.}
\label{ACPrattaumugalfextrm1}
\end{figure}

Now we would like to show our results  for the $A_{CP}$ of the process 
$\tau\rightarrow \mu\gamma$ in series of Figures \ref{ACPalftaumugsinalf01}
-\ref{ACPrattaumugalfextrm1}. Fig. \ref{ACPalftaumugsinalf01} is devoted to 
the new quantity $|\rho|$ dependence of $A_{CP}$ for 
$\bar{\xi}^E_{N,\tau\tau}=100\, GeV$, $sin\theta_{\rho}=0.1$, 
$sin\theta_{\tau\mu}=0.5$,  $m_{h^0}=85\, GeV$ and $m_{A^0}=95\, GeV$. Here, 
$A_{CP}$ is restricted between solid lines for $sin\theta_{\tau\tau}=0.1$, 
dashed lines for $sin\theta_{\tau\tau}=0.5$ and small dashed lines for 
$sin\theta_{\tau\tau}=0.8$. Note that the upper and lower bounds for 
$A_{CP}$ is due to the constraint of $\bar{\xi}^E_{N,\tau\mu}$ coming from 
the experimental limits of $\mu$-lepton EDM. $A_{CP}$ for this process is 
at the order of the magnitude of $10^{-3}$. However, the increasing values 
of $sin\theta_{\rho}$ enhances it almost one order as seen in Figures 
\ref{ACPalftaumugsinalf05} and \ref{ACPalftaumugsinalf08}, where they 
corresponds to $sin\theta_{\rho}$ values $0.5$ and $0.8$ respectively. 
The strong sensitivity of $A_{CP}$ to the parameter $sin\,\theta_{\rho}$ 
is shown in Fig. \ref{ACPsinalftaumugalfextrm1}.  In this figure, we 
present $sin\,\theta_{\rho}$ dependence of $A_{CP}$  for $|\rho|=0.1\, 
(GeV^2)$, $sin\theta_{\tau \mu}=0.5$, $m_{h^0}=85\, GeV$ and $m_{A^0}=95\, 
GeV$. Here, $A_{CP}$ is restricted between solid lines for $sin\theta_
{\tau\tau}=0.1$, dashed lines for $sin\theta_{\tau\tau}=0.5$ and small 
dashed lines for $sin\theta_{\tau\tau}=0.8$. These figures show that the 
restriction region for $A_{CP}$ becomes broader with increasing values of  
$sin\theta_{\tau\tau}$. The same behaviour appears when $sin\theta_{\tau\mu}$ 
increases also.

Finally, we study the mass ratio $m_{h^0}/m_{A^0}$ dependence of $A_{CP}$ 
for the fixed values of $sin\,\theta_{\tau \mu}=sin\,\theta_{\tau \tau}=
0.5$, $|\rho|=0.1\,(GeV^2)$ and $m_{h^0}=85\, GeV$ in Fig. 
\ref{ACPrattaumugalfextrm1}. Here $A_{CP}$ is restricted between solid lines 
for $sin\,\theta_{\rho}=0.1$, dashed lines for $sin\theta_{\rho}=0.5$ 
and small dashed lines for $sin\theta_{\rho}=0.8$.
It is observed that $A_{CP}$ increases when the masses of neutral Higgs 
bosons are near to degeneracy. This enhancement is large for large values 
of $sin\theta_{\rho}$.

As a result, it is possible to get a measurable $A_{CP}$ for the LFV 
interactions $\mu\rightarrow e\gamma$ and $\tau\rightarrow \mu\gamma$ if 
there exists an additional small and complex contribution coming from 
the physics beyond the model III. Here, the complexity of the new 
contribution should not be due to the Yukawa type couplings, but comes 
from radiative corrections. With the reliable experimental measurements 
of such asymmetries, it would be possible to test the new contributions 
and the corresponding physics. 
\section{Acknowledgement} This work was supported by Turkish Academy of
Sciences (TUBA-GEBIP).

\newpage
\newpage

\begin{thebibliography}{1}
%
\bibitem{Brooks} M. L. Brooks et. al., MEGA Collaboration, 
{\it Phys. Rev. Lett.} {\bf 83}, 1521 (1999).
%
\bibitem{Ahmed} S. Ahmed et.al., CLEO Collaboration,  
{\it Phys. Rev.} {\bf D61}, 071101 (2000).
%
\bibitem{Chang} D. Chang, W. S. Hou and W. Y. Keung, 
{\it Phys. Rev.} {\bf D48}, 217 (1993).
%
\bibitem{Diaz} R. Diaz, R. Martinez and J-Alexis Rodriguez, 
hep-ph/0010149 (2000).
%
\bibitem{Barbieri1} R. Barbieri and L. J. Hall, 
{\it Phys. Lett.} {\bf B338}, 212 (1994).
%
\bibitem{Barbieri2} R. Barbieri, L. J. Hall and A. Strumia, 
{\it Nucl. Phys.} {\bf B445}, 219 (1995).
%
\bibitem{Barbieri3} R. Barbieri, L. J. Hall and A. Strumia, 
{\it Nucl. Phys.} {\bf B449}, 437 (1995).
%
\bibitem{Ciafaloni} P. Ciafaloni, A. Romanino and A. Strumia, 
IFUP-YH-42-95.
%
\bibitem{Duong} T. V. Duong, B. Dutta and E. Keith, 
{\it Phys. Lett.} {\bf B378}, 128 (1996).
%
\bibitem{Couture} G. Couture, et. al., 
{\it Eur. Phys. J.} {\bf C7}, 135 (1999).
%
\bibitem{Okada} Y. Okada, K. Okumara and Y. Shimizu, 
{\it Phys. Rev.} {\bf D61}, 094001 (2000).
%
\bibitem{Czarnecki} A. Czarnecki and E. J. Jankowski, hep-ph/0106237. 
%
\bibitem{Lavignac} S. Lavignac, I Masina and C. A. Savoy, 
hep-ph/0106245. 
%
\bibitem{IltanLFV} E. Iltan,
{\it Phys. Rev.} {\bf D64}, 013013 (2001).
%
\bibitem{Sher} T. P. Cheng and M. Sher, {\it Phy. Rev.} {\bf D35}, 3383 
(1987). 
%
\bibitem{Abdullah} K. Abdullah et.al,  {\it Phys. Rev. Lett.} {\bf 65}, 
(1990) 2347. 
%
\end{thebibliography}
\end{document}